\documentclass[12pt]{article}
\usepackage[margin=2cm]{geometry}
\usepackage{comment}
\usepackage{amsmath,amssymb,extarrows,mathtools,graphicx,subfigure,setspace}
\usepackage{cite}
\usepackage{slashed}
\usepackage{color}
\usepackage{tikz}
\usepackage{fancyhdr}
\usepackage{cleveref}
\usetikzlibrary{decorations.pathmorphing,calc}
\makeatother

\newcommand{\ud}{\,\mathrm{d}}

\newcommand{\be}{\begin{equation}}
\newcommand{\bea}{\begin{eqnarray}}
\newcommand{\eea}{\end{eqnarray}}
\newcommand{\ba}{\begin{array}}
\newcommand{\ea}{\end{array}}
\newcommand{\ee}{\end{equation}}
\newcommand{\bes}{\begin{equation*}}
\newcommand{\beas}{\begin{eqnarray*}}
\newcommand{\eeas}{\end{eqnarray*}}
\newcommand{\bas}{\begin{array*}}
\newcommand{\eas}{\end{array*}}
\newcommand{\ees}{\end{equation*}}

\numberwithin{equation}{section}
\begin{document}
	\onehalfspacing
	\noindent

	\begin{titlepage}
		\vspace{10mm}
		\begin{flushright}
		\end{flushright}
		
		\vspace*{20mm}
		\begin{center}
			
			{\Large {\bf Holographic complexity of Born-Infeld gravity}\\
			}
			\vspace*{15mm}
			{Hamid R. Bakhtiarizadeh ${}^{a,}$\footnote{bakhtiarizadeh@sirjantech.ac.ir}, Ghadir Jafari ${}^{b,}$\footnote{ghjafari@ipm.ir} }
			\\[5mm]
				{\it ${}^a$ Department of Physics, Sirjan University of Technology, Sirjan, Iran\\
				${}^b$ School of Particles and Accelerators,
				Institute for Research in Fundamental Sciences (IPM)\\
				P.O. Box 19395-5531, Tehran, Iran	
			}
			
			\vspace*{1mm}

			\vspace*{1cm}

			\vspace*{1cm}
		\end{center}
		
		\begin{abstract}
We investigate the duality conjecture ``Complexity=Action" (CA) for  Born-Infeld (BI) gravity model  and derive the growth rate of its action within the Wheeler-DeWitt (WDW) patch, which is believed to be dual to the growth rate of quantum complexity of holographic boundary state. In order to find the correct on-shell action, we use direct variational procedure to find proper boundary and corner terms. We  find out that the late-time behavior of holographic complexity is the well-known two times of energy, as expected.

		\end{abstract}
		
	\end{titlepage}

\section{Introduction}

The AdS/CFT proposal \cite{Maldacena:1997re} in recent years inclined to relate quantities in quantum information theory to those of (quantum) gravitational theory. The first example of such relation is the Ryu-Takayanagi
proposal  which  provides  a  geometrical  realization  of  entanglement  entropy  in  a  dual  CFT \cite{Ryu:2006bv}.
 Another example is CA proposal \cite{Brown:2015lvg,Stanford:2014jda}, which relates the quantum computational complexity of a boundary state to the on-shell action on a bulk region named as the WDW patch
\begin{equation}\label{CA}
{\cal C}(\Sigma)=\frac{I_{\text{WDW}}}{\pi \hbar}~.
\end{equation}
 Here, the time slice $ \Sigma $ is defined as the intersection of any Cauchy surface in the bulk with the asymptotic boundary, where $ {\cal C} $ is to be evaluated. It is also conjectured that there is a bound on the growth rate of this holographic complexity according to \footnote{However we must point that as found in \cite{Carmi:2017jqz} this bound is violated for Einstein gravity.}:
 \begin{equation}\label{Cbound}
\frac{ d\mathcal{C}}{dt}\le \frac{2M}{\pi \hbar}~,
 \end{equation}
which is thought to be holographic dual the Lloyd’s bound on the quantum complexity. 
It also was found in \cite{Brown:2015lvg,Stanford:2014jda} that at late times, this bound is saturated
\begin{equation}\label{AG}
\frac{d\mathcal{C}}{dt}= \frac{2M}{\pi \hbar}~,
\end{equation}
for an AdS-Schwarzschild black hole, where $ M $ is the total mass-energy of the spacetime, and $ t $ stands for one of the boundary times.

One may question the validity of the above proposal in the presence of higher-derivative theories. This question has been investigated for $ f(R) $ and critical gravities \cite{Alishahiha:2017hwg}, Gauss-Bonnet and Lovelock theories \cite{Cai:2016xho,Cano:2018aqi}. In this article we are going to pursue this question for a model of higher-derivative theories that includes infinite number of derivatives in general dimensions. This model known as Born-Infield gravity and was first proposed in \cite{Gullu:2010pc} as an extension of New Massive Gravity (NMG) in three dimension. The importance of higher-derivative theories in the AdS/CFT prospective is that higher-derivative or stringy corrections in bulk space correspond to considering smaller value for the 't Hooft coupling in the boundary side \emph{i.e.,} $ \lambda\propto\frac{L^4}{\alpha'^2} $. 
This suggest that investigating infinite theories in spacetime means considering field theory at weak couplings.  

To study the holographic complexity, one needs to completely determine appropriate boundary and corner terms in the gravitational action to obtain correct on-shell value of the action\cite{Lehner:2016vdi}. In this regard the main obstacle in the way of studying higher-derivative theories is finding these boundary terms (see \emph{e.g.} \cite{Dyer:2008hb}). We note that these boundary terms are needed in order to have a well-posed variational principle, so that these terms could be found by direct variation of gravitational action and checking consistency with Dirichlet boundary condition \cite{Padmanabhan:2014lwa}. This procedure is not applicable to many higher-derivative theories. In this article we found that, this method is applicable for an special model of higher-derivative theories in Einstein backgrounds. As a result, we found appropriate boundary terms to make variational principle well-posed.  Interesting property that we observed is that the boundary terms are proportional to the standard boundary terms for Einstein-Hilbert theory. Using this fact and evaluating the on-shell bulk Lagrangian enabled us to find out that the whole on-shell action in this theory is proportional to Einstein-Hilbert one. Therefore, we conclude that the late-time behavior is scaled by the same factor with respect to Einstein-Hilbert theory. At the end we show that the same factor appears when comparing of energy in these theories. This means that the late-time complexity growth is the same two times of energy. 

The outline of the paper is arranged as follows. In the next section we first introduce the BI gravity action in general dimensions. Then we use some manipulation to obtain the equations of motion, and investigate static spherical symmetric black hole solutions. In Sec. \ref{Btensor} we find proper boundary terms needed to calculate the on-shell action in WDW patch that includes null and joint terms. Finally, in Sec. \ref{complexity} we use these results to calculate action growth rate for black holes in BI gravity. 


\section{BI gravity}

The BI gravity, where first was introduced in \cite{Gullu:2010pc} as a generalization of NMG in three dimensions, is described by the following action:
\begin{equation}\label{BIG}
I_{\rm BI}=-\frac{4m^2}{\kappa^2}\int d^dx\sqrt{-\det g}\bigg[\sqrt{\det(\mathbf{1}+\tfrac{\sigma}{m^2}g^{-1}G)}-\lambda\bigg]~,
\end{equation}
where $ G_{\mu\nu}=R_{\mu\nu}-\tfrac12 R g_{\mu\nu} $ is the Einstein tensor, $m^2$ is a positive definite dimension-full parameter, $\sigma=\pm1$ fixes the sign of the Einstein tensor in the first term, $\lambda$ is a parameter related to the cosmological constant and $\kappa$ is related to  the $ d $-dimensional Newton gravitational constant. Expanding this action in powers of the parameter $1/m^2$, which is practically derivative expansion, has the following feature
\begin{align}\label{NMG}
I=-\frac{4m^2}{\kappa^2}\int d^dx\sqrt{-\det g}\big[(1-\lambda) - \frac{(d-2) \sigma}{4 \mathit{m}^2}R - \frac{1}{4\mathit{m}^4} \bigl( R_{\alpha \beta } R^{\alpha \beta } - \tfrac{1}{8} (d^2- 6 d + 12) R^2\bigr)+{\cal O}(R^3)\big]~.
\end{align}
The first order expansion gives the Einstein-Hilbert (EH) action with an effective cosmological constant $\Lambda_\text{eff}=2m^2(1-\lambda)/\sigma(d-2)$. Up to the second order in three dimensions it reproduces the NMG action \cite{Bergshoeff:2009hq}, which is a non-chiral massive gravity in three dimension. It is also easy to show that expanding the BI action (\ref{BIG}) up to the third order reproduces the extension of NMG theory to fourth order computed in \cite{Sinha:2010ai}. However, here we are interested in features of this theory in general dimensions, as a model of higher derivative with infinite number of derivatives.  

To find the equations of motion, we use the following formula for a general matrix $ \mathbf{A}$,
\begin{equation}\label{delta1}
\delta(\sqrt{\det \mathbf{A}})=\tfrac12\sqrt{\det \mathbf{A}}\, \, \, \text{Tr}(\mathbf{A}^{-1}\delta\mathbf{A})~,
\end{equation}
where $ \mathbf{A}.\mathbf{A}^{-1}=\mathbf{1} $. 
Looking at action (\ref{BIG}), one can identify the matrix as $ \mathbf{A}^{\mu}{}_{\nu} =\delta^{\mu}{}_{\nu} +\tfrac{\sigma}{m^2}G^{\mu}{}_{\nu} $, for which the variation is given by $ \delta \mathbf{A}^{\mu}{}_{\nu} =\tfrac{\sigma}{m^2}\delta G^{\mu}{}_{\nu}  $.
To obtain the equations of motion, it is convenient to define a tensor
\begin{equation}\label{Vdef}
\mathcal{V}^{\mu}{}_{\nu}:=\sqrt{-\det(\mathbf{1}+\tfrac{\sigma}{m^2}g^{-1}G)}(\frac{1}{\mathbf{1}+\tfrac{\sigma}{m^2}g^{-1}G})^{\mu}{}_{\nu}~,
\end{equation}
and an operator,
\begin{align}\label{Pdef}
P_{\mu\nu\alpha\beta}:=&g_{\mu\nu}R_{\alpha\beta}-g_{\alpha\mu}g_{\beta\nu}(R+\square)\nonumber\\&+g_{\beta\mu}\nabla_{\alpha}\nabla_{\nu }+g_{\beta\nu}\nabla_{\alpha}\nabla_{\mu}-g_{\mu\nu}\nabla_{\alpha}\nabla_{\beta}
+g_{\mu\nu}g_{\alpha\beta}\square-g_{\alpha\beta}\nabla_{\mu}\nabla_{\nu}~.
\end{align}
By which we can express the variation of Einstein tensor as:
\begin{align}\label{deltaG}
\delta G_{\mu\nu}&=\tfrac12 P_{\mu\nu\alpha\beta}\delta g^{\alpha\beta}~.
\end{align}
Using the above  equations one may show that the  equations of motion have the following compact form,
\begin{equation}\label{EOM}
\frac{\sigma}{2m^2}P_{\mu\nu\alpha\beta}\mathcal{V}^{\alpha\beta}  +g_{\mu\nu}\bigg[\sqrt{-\det(\mathbf{1}+\tfrac{\sigma}{m^2}g^{-1}G)}-\lambda\bigg]=0~.
\end{equation}

In order to speculate the holographic complexity of this theory, we are interested in AdS-Schwartzshild black hole solutions of this theory. These solutions we can consider to be as: 
$ R_{\alpha\beta}=-\tfrac{d-1}{\ell^2}g_{\alpha\beta} $. Using this relation 
one can rewrite the tensor $ \mathcal{V}^{\mu\nu} $ as: 
\begin{equation}\label{Vads}
\mathcal{V}^{\mu\nu}=\bigl[1 +  \frac{(d-2) (d-1)  \sigma}{2 \ell^2\mathit{m}^2}\bigr]^{\frac{d-2}2}g^{\mu\nu}~. 
\end{equation}
As a result, the equation of motion \eqref{EOM} simplifies to
\begin{equation}\label{Eom2}
\bigl[1 + \frac{(d-2) (d-1)  \sigma}{2 \ell^2\mathit{m}^2}\bigr]^{\frac{d-2}2}-\lambda=0~.
\end{equation}
By solving this equation for $ \ell^2 $, one arrives at
\begin{equation}\label{ell}
\frac{1}{\ell^2}= \frac{2\sigma \mathit{m}^2 (\lambda^{\frac{2}{d-2}}-1) }{(d-2) (d-1)}~.
\end{equation}
This is the effective AdS radius due to the higher derivatives. So the line element for static black holes in this theory is given by:
\begin{equation}\label{metric}
ds^2=-f(r)dt^2+\frac{1}{f(r)}dr^2+r^2d\Sigma_{d-2,\kappa}^2,
\end{equation}
where $ f(r) =\kappa+\tfrac{r^2}{\ell^2}-\frac{\omega^{d-3}}{r^{d-3}}$, and $ \kappa $ is the curvature of the $ (d-2)  $-dimensional line element $ d\Sigma_{d-2}^2 $ and is $ \kappa=\{1,0,-1\} $ for spherical, planar, and hyperbolic horizon geometries, respectively. As a result AdS-Schwartzshild black holes are solutions of this theory provided that the asymptotic AdS radius is given by \cref{ell}.


\section{Boundary and joint terms  for BI gravity}\label{Btensor}
In the previous section we analyzed the BI theory, its equations of motion and AdS solution. When one vary an action in order to find the equations of motion, usually deals with a surface term that by supposing appropriate boundary or fall-off conditions leads to a consistent variational problem. In the case of gravitational action these surface terms contain metric and its normal derivative, and fixing both on the boundary is not consistent with equations of motion. In these cases usually variational principle in restored by adding some terms to the action on the boundary of  spacetime. The appropriate boundary terms can be found by analyzing the surface  integral. In the following we analyze the surface terms on variations of BI action in order to find its boundary terms in different kinds of boundaries.
\subsection{Spacelike and timelike boundaries}
In finding the equations of motion \eqref{EOM} we have used integration by part. The total derivative of this integration leads to the term $ \int_{\partial\mathcal{M}}\sqrt{-h}\ n_{\alpha}J^{\alpha}$ on the boundary, where the expression for $ J^\alpha $ is as following 
\begin{align}
	J_{\alpha}&=\sigma\big[- \delta \mathit{g}^{\gamma}{}_{\gamma} \nabla_{\alpha}\mathcal{V}^{\beta}{}_{\beta} + \delta \mathit{g}_{\beta \gamma} \nabla_{\alpha}\mathcal{V}^{\beta \gamma} -  \mathcal{V}^{\beta \gamma} \nabla_{\alpha}\delta \mathit{g}_{\beta \gamma} + \mathcal{V}^{\beta}{}_{\beta} \nabla_{\alpha}\delta \mathit{g}^{\gamma}{}_{\gamma} + \delta \mathit{g}^{\gamma}{}_{\gamma} \nabla_{\beta}\mathcal{V}_{\alpha}{}^{\beta} + \mathcal{V}^{\beta \gamma} \nabla_{\beta}\delta \mathit{g}_{\alpha \gamma} \nonumber\\&-  \mathcal{V}^{\beta}{}_{\alpha} \nabla_{\beta}\delta \mathit{g}^{\gamma}{}_{\gamma} + \mathcal{V}^{\beta \gamma} \nabla_{\gamma}\delta \mathit{g}_{\alpha \beta} -  \mathcal{V}^{\beta}{}_{\beta} \nabla_{\gamma}\delta \mathit{g}_{\alpha}{}^{\gamma} -  \delta \mathit{g}_{\beta \gamma} \nabla^{\gamma}\mathcal{V}_{\alpha}{}^{\beta} -  \delta \mathit{g}_{\beta \gamma} \nabla^{\gamma}\mathcal{V}^{\beta}{}_{\alpha} + \delta \mathit{g}_{\alpha \gamma} \nabla^{\gamma}\mathcal{V}^{\beta}{}_{\beta}\big]~.
\end{align}
Although the above expression seems complicated and finding proper boundary terms for general metric background is impossible, but for an Einstein space background and using \cref{Vads}  it simplifies to:
\begin{equation}\label{BT2}
	J_{\alpha}=(d-2)\bigl[1 +  \frac{(d-2) (d-1)  \sigma}{2 \ell^2\mathit{m}^2}\bigr]^{\frac{d-2}2}\sigma (\nabla_{\beta}\delta \mathit{g}_{\alpha}{}^{\beta}-\nabla_{\alpha}\delta \mathit{g}^{\beta}{}_{\beta})~,
\end{equation}
which upon using the equations of motion \eqref{Eom2} it becomes:
\begin{equation}\label{BT3}
J_{\alpha}=(d-2)\lambda\sigma (\nabla_{\beta}\delta \mathit{g}_{\alpha}{}^{\beta}-\nabla_{\alpha}\delta \mathit{g}^{\beta}{}_{\beta})~.
\end{equation}
This surface integral is nothing but the surface integral of EH action under variations by a factor $ (d-2)\sigma\lambda $. It is well-known that using these terms one can obtain the Gibbons–Hawking–York (GHY) term and the Brown-York (BY) stress tensor, (see \emph{e.g.} \cite{Padmanabhan:2014lwa}). In fact if we have a spacelike or timelike boundary, it can be shown that:
\begin{align}\label{var1}
\int_{\mathcal{M}} \ud^{d}x \sqrt{-g} \nabla^{\alpha}\left(\nabla_{\alpha}\delta g_{\beta}^{\beta}-\nabla_{\beta}\delta g_{\alpha}^{\beta}\right)=&2\int_{\partial\mathcal{M}}  \ud^{d-1}x \delta(\sqrt{|h|}K)+2\int_{C}\ud^{d-2} x \,\delta(\sqrt{|q|}\vartheta)\nonumber\\&
+\int_{\partial\mathcal{M}}  \ud^{d-1}x \sqrt{|h|} (K_{ab}-h_{ab}K)\delta h^{ab}~,
\end{align}
where $ h_{ab} $ is induced metric in the boundary $ \partial\mathcal{M} $, and $ K_{ab} $ is its extrinsic curvature. $ \vartheta $ is angle or rapidity defined in regions where two segments of boundary reach each other non-smoothly and are shown by $C$.   The second line of this expression vanishes for example by choosing the Dirichlet boundary condition \emph{i.e.,} $ \delta h_{ab}=0 $.  
As a result this equation says that in order to have an action with a well-posed variational principle, we must subtract a boundary GHY term from the action on timelike or spacelike boundaries. Furthermore, if the boundary is not smooth or two segments of boundary join together, a so called ``joint terms" on these co-dimension surfaces are needed in order to variational principle become well-posed \cite{Hayward:1993my} (for details one also can refer to \cite{Lehner:2016vdi,Aghapour:2018icu}). The second line can be used to find the BY stress  tensor. If $ h_{ab} $ be the induced metric on a timelike boundary, according to Brown and York \cite{Brown:1992br},
differentiation with respect to it gives the quasilocal definition for energy of gravitational field.  We will use this expression also to find correct energy of black holes in this theory in  section \ref{complexity}. Therefore we conclude that appropriate boundary term for BI theory on timelike or spacelike boundary is given by:
\begin{equation}\label{BIboundary}
S_b=-2 (d-2)\lambda\sigma\int_{\partial\mathcal{M}}  \ud^{d-1}x \sqrt{|h|}K-2(d-2)\lambda\sigma\int_{C}\ud^{d-2} x \,\sqrt{|q|}\vartheta.
\end{equation} 
\subsection{Null boundaries}
Similar analysis can be done for the case of null boundaries. Decomposing the surface term on null boundaries gives us corresponding term for a well-posed variational principle\cite{Lehner:2016vdi,Parattu:2015gga} and a similar stress tensor is defined on such boundaries \cite{Jafari:2019bpw}. The corresponding expression in the case of null boundaries is\cite{Aghapour:2018icu}:
\begin{align}\label{varnull}
&\int d^{d}x \sqrt{-g} \nabla^{\alpha}\left(\nabla_{\alpha}\delta g_{\beta}^{\beta}-\nabla_{\beta}\delta g_{\alpha}^{\beta}\right)=2\int\!\!\! \ud^{d-1} x \, \,\delta\left(\sqrt q [(\Theta + \kappa)] \right)+2\int\!\!\! \ud^{d-2} x \, \delta\left( \sqrt q \,\ln A\right)\nonumber \\
&+\int\!\!\! \ud^{d-1} x \,  \sqrt q\left(\left[\Theta^{ab} -q^{ab}\,(\Theta + \kappa)\right]\delta q_{ab}+ 2\,\omega^a\,\delta \beta_{a}  -2 \Xi\, \delta B\right)+\int\!\!\! \ud^{d-2} x\sqrt q (\ln A q^{ab})\delta q_{ab}~.  	
\end{align} 
Here the null boundary is specified by $ \phi=const. $, and its normal given by $ \ell_{\alpha}=A \nabla_{\alpha}\phi\ $. In order to describe a null hypersurface, in addition to  $ \ell_{\alpha}$ we need a second auxiliary null vector $ k_a $ which is transverse to the null boundary  and is normalized according to the relation $ \ell^ak_a=-1 $. The quantities $ \Theta_{ab},\Xi_{ab},\omega_a $ and $ \kappa $ are geometric objects which represent extrinsic geometry of the hypersurface and are defined as:  

\begin{align}
\Theta_{ab} = q^c{}_a\,q^d{}_b\,\nabla_a\ell_b \ , \quad \Xi_{ab} = q^c{}_a\,q^d{}_b\,\nabla_a\,k_b,\quad \omega_a = - q^c{}_a\,k^b\,\nabla_c\ell_b , \quad \kappa = -\ell^a\,k^b\,\nabla_a\ell_b~.
\end{align}
Here also $ q^a_b $ is a projection on a codimension-two surface on the null boundary defined by $ q^a_b=\delta_a^b +\ell^ak_b+\ell_bk^a $. Furthermore, we point that $ \{\delta q_{ab},\delta\beta_{a},\delta B\} $ are variation of metric components along the boundary, and accordingly can be fixed by choosing a Dirichlet boundary condition\cite{Jafari:2019bpw}. As a result second line in \cref{varnull} vanishes due to the boundary condition and we are left with the first line. In fact the terms in the first line of \cref{varnull} indicate those terms that are needed to be subtracted on a null boundary to have a well-posed variational principle. So in order to have a well-defined variational principle, we must have the following terms on the null boundary for BI theory:
\begin{equation}\label{BInull}
S_b=-2(d-2)\lambda\int_{\partial\mathcal{M}} \!\!\! \ud^{d-1} x \, \,\sqrt q [(\Theta + \kappa)]-2(d-2)\lambda\int_{\mathcal{C}}\!\!\! \ud^{d-2} x \,  \sqrt q \,\ln A~.
\end{equation}

Finally the whole well-posed action in BI theory  for AdS-Schwartzshild  background is given by:
\begin{align}\label{wellposedaction}
I=&-\frac{4m^2}{\kappa^2}\int\ud^{d}x\sqrt{-\det g}\bigg[\sqrt{\det(\mathbf{1}+\tfrac{\sigma}{m^2}g^{-1}G)}-\lambda\bigg]\nonumber\\&+\frac{(d-2)\lambda \sigma}{\kappa^2} \big(\int \ud^{d-1}xK_{t}\hspace{1mm}+\int\ud^{d-1}x K_{s}\hspace{1mm} + \int\ud^{d-1}x K_{n}\hspace{1mm}
+\int \ud^{d-2}x a\hspace{1mm}\big)~,
\end{align}
where $ K_{s},~K_{t},~K_{n} $ respectively denote boundary terms for spacelike, timelike and null in Einstein theory, and $ a $ stands for joint terms. We also must note that similar to the Einstein theory, the boundary terms in \eqref{BInull} have an ambiguity under reparametrization of null generators. This ambiguity for Einstein theory has been discussed in detail in \cite{Lehner:2016vdi,Carmi:2017jqz}. To overcome this ambiguity, the usual method is to add a further \textit{counterterm } to the action on the null boundary. For Einstein theory this counterterm is given by \cite{Lehner:2016vdi}: 
\begin{equation}\label{counter}
S_{ct}=\int_{\partial\mathcal{M}} \!\!\! \ud^{d-1} x \, \,\sqrt q \Theta\ln(l_{ct}\Theta)~.
\end{equation}
Here we argue that the corresponding term for BI theory in this paper is also proportional to the above term. In order to deduce that, we note that the algorithm for finding such terms, as presented in \cite{Lehner:2016vdi}, is to consider the effect of infinitesimal reparametrization of null generators as: $ \ell_{\alpha}\to\ell_{\alpha}+\beta(x)\ell_{\alpha} $ and $ k_{\alpha}\to k_{\alpha}-\beta(x)k_{\alpha} $. Under these reparametrizations $ S_{b} $ changes according to $ S_{b}\to S_{b}+\delta S_{b} $ and counterterms are found such that their transformations cancel the $ \delta S_{b} $ term. So we can easily see that since $ S_{b} $ in our theory is proportional to the Einstein theory, the corresponding counterterm is also  the above term with the same factor of proportionality.

\section{Holographic complexity}\label{complexity}
\begin{figure}
	\begin{center}
	\includegraphics[scale=.55]{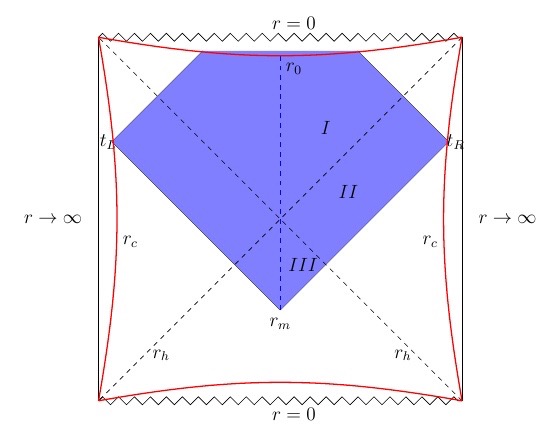}
\caption{Penrose diagram and WDW patch of an AdS-Schwarzschild black hole}\label{wdw}
	\end{center}
\end{figure}

Having found necessary boundary terms in the theory, we can now go for the calculation of the on-shell action. According to CA holographic complexity proposal the value of the on-shell action in a WDW patch of spacetime correspond to complexity of the dual theory.  WDW patch of AdS-Schwarzschild black hole spacetime is shown by blue color in the Penrose diagram Fig. \ref{wdw}. As we see the region contains several null segments and spacelike boundary as a cut off near the singularity as well as some joint of null segments.
Using the relations \eqref{Eom2} and \eqref{ell}, and after some algebra one can find the following value for the on-shell bulk Lagrangian:
\begin{equation}\label{onshelllag}
\mathcal{L}_{on-shell}=-2\lambda \sigma (d-2) (d-1) \frac{1}{\ell^2}~.
\end{equation}
If we compare this result with the on-shell action in EH theory, we observe that
\begin{equation}\label{onshell}
\mathcal{L}_{on-shell}=\sigma\lambda (d-2)\mathcal{L}_{EH}~.
\end{equation}
Note that the value of on-shell action in EH theory is $ -2(d-1)/\ell^2 $. So the whole on-shell bulk Lagrangian comparing to the Einstein theory has an extra factor $\sigma\lambda (d-2)$. Interestingly enough this is the same factor we have found for boundary terms comparing to the Einstein theory. This important property helps us to conclude the following relation for holographic complexity in BI theory:
\begin{equation}\label{complexityBI}
\mathcal{C}_{BI}=\sigma\lambda (d-2) \mathcal{C}_{EH}.
\end{equation} 
According to this relation, one may  deduce that the late-time complexity rate to be:
\begin{equation}\label{CRate}
\frac{d\mathcal{C}}{dt}= \sigma\lambda (d-2) \frac{2M}{\pi \hbar}~.
\end{equation}
Although this relation is mathematically true, but the point is that the parameter $ M $ appearing in the metric \eqref{metric} is the total energy of  black hole spacetime in Einstein gravity, but it is not total energy in BI theory.  
The energy of a black hole system can be obtained using BY stress tensor when properly regularized \cite{Balasubramanian:1999re}. 
In the other words, according to this method, the energy is calculated by the relation
\begin{equation}\label{energy}
E=\int_S\ud^{d-2}x\sqrt{q}T_{ij}u^iu^j~.
\end{equation}
The advantage of our procedure is that according to relation \eqref{var1} it shows that the stress tensor is also factorized by $ (d-2)\sigma \lambda $ and as a result the energy of black holes in BI theory is a factor of energy of same solution in Einstein theory. Therefore, at the late time we recover the same relation for complexity growth rate, namely,
\begin{equation}\label{key}
\frac{d\mathcal{C}}{dt}= \frac{2E}{\pi \hbar}~.
\end{equation}
This shows that the late-time growth rate is the same two times of energy in BI theory.
\section{Conclusion}
In this paper we examined the holographic complexity proposal for a model of higher derivative with infinite number of terms when expanded in terms of curvature terms. We have four proper boundary terms for this theory in order to have a well-posed variational principle. We have shown that on-shell bulk and boundary terms scale with the same factor when  compared with Einstein theory. Our results also show that the late-time rate of holographic complexity is the same two times of energy for AdS-Schwarzschild black holes. This result confirms universality of holographic complexity proposal, an infinite derivative theory in gravity side correspond to a field theory with small 't Hooft coupling.  
\section{Acknowledgement}

The work of H.B. has been fiancially supported by the research deputy of Sirjan University of Technology. The tensor calculations in this paper has been carried out by the Mathematica package ``xTras" \cite{Nutma:2013zea}.

\appendix

\end{document}